\documentclass[runningheads,a4paper]{llncs}
\usepackage{url}
\usepackage{amsmath}
\usepackage{graphicx}

\usepackage{MnSymbol}
\usepackage{hyperref}

\urldef{\mailsa}\path|ameer.tawfik@gmail.com|
\urldef{\mailsb}\path|naomie@utm.my|   
\newcommand{\keywords}[1]{\par\addvspace\baselineskip
\noindent\keywordname\enspace\ignorespaces#1}
\begin{document}

\mainmatter
\title{Adapting Voting Techniques for Online Forum Thread Retrieval}

\author{Ameer Tawfik Albaham, 
        Naomie Salim
        }
        
\institute{Faculty of Computer Science and Information Systems,\\
Universiti Teknologi Malaysia, Skudai, Johor,Malaysia\\
\mailsa,
\mailsb
}
\maketitle
\begin{abstract}
Online forums or message boards are rich knowledge-based
communities. In these communities, thread retrieval is an essential tool
facilitating information access. However, the issue on thread search is how
to combine evidence from text units(messages) to estimate thread relevance. 
In this paper, we first rank a list of messages, then we score threads
by aggregating their ranked messages' scores. To aggregate the message
scores, we adopt several voting techniques that have been applied in
ranking aggregates tasks such as blog distillation and expert finding.
The experimental result shows that many voting techniques should be
preferred over a baseline that treats a thread as a concatenation of its message texts.

\end{abstract}

\section{Introduction}
Online forums are virtual places(communities) that facilitate seeking and sharing 
knowledge through in depth discussions. A user starts a discussion through
posting an initial message, then other users read the initial message and answer
it through reply messages. The initial message and its replies form a threaded
discussion(thread).
One challenge in accessing information in forums is information overload.
Thread retrieval is one way to tackle it. However, the actual contents are not
the threads but the messages. Therefore, given a query, a retrieval system must
infer the thread relevance using the message text. In that aspect, thread retrieval
resembles ranking aggregates tasks such as blog feed retrieval\cite{seo2008,elsas2008,macdblog} 
and expert finding\cite{macdexpert}. 
In these tasks, given a query, the objective is to rank aggregates (blogs, experts) by leveraging associated text units (blogs' postings, experts' writings)\cite{macd2011}. 
An analogy between ranking aggregates and thread retrieval is that
threads are the aggregates, and messages are the associated texts or documents.

Voting techniques performed well in ranking aggregates tasks  \cite{macdexpert,macdblog,macd2011}. 
However, the effectiveness of each voting technique varies between tasks and datasets \cite{macd2011}.
In addition to that, threads have a conversational structure that does not exist
in other ranking aggregates contexts. In threads, the meaning of a message is
fully understood within its discussion context. Furthermore, messages are mostly
replies, hence they tend to be shorter than blogs' postings and experts' writings.
In other words, that might alter the performance of voting techniques.
In this paper, we review several voting methods and investigate their performance on thread retrieval.

\section{Voting in Thread Retrieval}

Voting techniques were first proposed by \cite{macdexpert} to the expert finding task. 
In voting techniques, we first rank a list of documents (e.g. expert's writings) based on
their relevance to the given query. Then, we rank aggregates(e.g., the experts) 
based on their scores obtained from fusing their ranked documents' scores or
ranks.
Similarly, in this work, given a query $Q = \{q_1,q_2,...,q_n\}$, we first rank a list
of messages $R_Q$ with respect to $Q$. 
Then, we score threads by aggregating their ranked messages' scores or ranks.
In addition, threads are ranked based on their obtained aggregated scores in a descending order.

In estimating the relevance between the query Q and a message M, we employ
the query language model\cite{ponte1998} assuming term independence, uniform probability
distribution for M and Dirichlet smoothing as follows\cite{zhai2004}:

\begin{equation}
\label{eq-qlm}
  P(Q|M)=
  \prod_{q\in Q}
  	\left(
  		\frac
  		{n(q,M)+\mu P(q|C)}
  		{|M|+\mu}
  		\right)
  		^{n(q,Q)}
\end{equation}
where $q$ is a query term, $\mu$ is the smoothing parameter. $n(q,M)$ and $n(q,Q)$
are the term frequencies of $q$ in $M$ and $Q$ respectively, $|M|$ is the number of
tokens in $M$, and $P(q|C)$ is the collection language model.
The outputs of $P(Q|M)$ and $P(Q|C)$ are probabilistic values.

To rank threads, the twelve aggregation methods proposed by \cite{macdexpert} are adapted:
Votes, Reciprocal Rank(RR), BordaFuse, CombMIN, CombMAX, CombMED,
CombSUM, CombANZ, CombMNZ, expCombSUM, expCombANZ and expCombMNZ. 
In addition to these methods, this study uses “CombGNZ”--- the geometric mean of the relevance scores. 
We use this method because it is the aggregation method employed by \cite{elsas2009,seo2011}.

In these methods, the relevance between a thread T and Q, $rel(T,Q)$, 
is the score obtained through the aggregation of all $T$'s ranked messages $R_T$
as shown below:
\begin{equation}
rel_{\text{Votes}}(Q,T) = |R_T|
\end{equation}

\begin{equation}
rel_{\text{RR}}(Q,T) = \sum_{M \in R_T} \frac{1}{rank(Q,M)}
\end{equation}

\begin{equation}
rel_{\text{BordaFuse}}(Q,T) = \sum_{M \in R_T} |R_Q| -− rank(Q,M)
\end{equation}

\begin{equation}
rel_{\text{CombMIN}}(Q,T) = MIN_{M \in R_T} P(Q|M)
\end{equation}

\begin{equation}
rel_{\text{CombMAX}}(Q,T) = MAX_{M \in R_T} P(Q|M)
\end{equation}

\begin{equation}
rel_{\text{CombMED}}(Q,T) = Median_{M \in R_T} P(Q|M)
\end{equation}

\begin{equation}
rel_{\text{CombSUM}}(Q,T) = \sum_{M \in R_T} P(Q|M)
\end{equation}

\begin{equation}
rel_{\text{CombANZ}}(Q,T) = \frac{1}{R_T} \times \sum_{M \in R_T} P(Q|M)
\end{equation}

\begin{equation}
rel_{\text{CombGNZ}}(Q,T) = \left(\prod_{M \in R_T} P(Q M) \right)^\frac{1}{|R_T|}
\end{equation}

\begin{equation}
rel_{\text{CombMNZ}}(Q,T) = |R_T| \times \sum_{M \in R_T} P(Q|M)
\end{equation}

\begin{equation}
rel_{\text{expCombSUM}}(Q,T) = \sum_{M \in R_T} \exp(P(Q|M))
\end{equation}

\begin{equation}
rel_{\text{expCombANZ}}(Q,T) = \frac{1}{|R_T|} \times \sum_{M \in R_T} \exp(P(Q|M))
\end{equation}

\begin{equation}
rel_{\text{expCombMNZ}}(Q,T) = |R_T| \times \sum_{M \in R_T} \exp(P(Q|M))
\end{equation}
where $rank(Q,M)$ is the rank of the message $M$ in $R_Q$, $|R_Q|$
is the size of $R_Q$, and $|R_T|$ is the number of $T$'s ranked messages.

As an illustrative example, let $R_Q = \{M_1,M_2,M_3,M_4,M_5,M_6\}$ denote a list of ranked messages,
where there are 3 threads associated with these messages $T_1$,$T_2$ and $T_3$; and, $M_1$ belongs to $T_1$,
$M_2$ and $M_3$ belong to $T_2$ and $M_4$, $M_5$ and $M_6$ belong to $T_3$.
In addition, let the relevance scores between the user query and these messages assigned by query language relevance model to be
0.06, 0.05, 0.04, 0.03, 0.02 and 0.01 respectively, whereas the ranks of these messages are
1,2,3,4,5,6. Then, we calculate the relevance between the given query $Q$ and the thread $T_6$ using the Votes, the CombSUM and the BordaFuse aggregation methods as follows:
$rel_{\text{Votes}}(Q,T_6) = |R_{T_6}| = 3$, 
$rel_{\text{CombSUM}}(Q,T_6) = P(Q|M_4) + P(Q|M_5) + P(Q|M_6) = 0.04 + 0.05 + 0.06 = 0.16$,  
$rel_{\text{BordaFuse}}(Q,T_6) = 6 - rank(Q,M_4) + 6 - rank(Q,M_5) + 6 - rank(Q,M_6) = 2 + 1 + 0 = 16$.

\section{Related Studies}

The voting techniques approach to the ranking aggregates tasks are inspired by works on data fusion (meta search)\cite{df-fox,ogilvie2003,df-bordafuse-2001}. 
A meta search algorithm aims to combine several ranked lists of documents into a unified list \cite{df-bordafuse-2001}. 
These ranked lists are generated by various retrieval methods. 
The essences of the data fusion are two folds\cite{spoerri2008}. 
First, the more retrieval methods retrieve a particular document, the more the document is expected to be relevant to the user query.
Second, a document that is ranked at top ranking positions by many retrieval methods might be more relevant than a one that was found at the bottom of several ranked lists. 
Data fusion can be categorized into score based and rank based aggregation methods. 
The score based methods --- such as \cite{df-fox}'s CombMAX, CombMIN, CombMED and CombSUM methods, use the relevance scores of documents, whereas the rank based methods, \cite{df-bordafuse-2001}, utilize the ranking positions of these documents on the ranked lists. 

\cite{macdexpert,macdblog} approached the problem of ranking aggregates as a data fusion problem:
each document is an evidence about its parent aggregate's relevance to the query. 
Generally, the the voting approach was found to be statistically superior to baseline methods \cite{macdexpert,macdblog}.
However, the performance of each voting technique was not consistent across tasks\cite{macd2011}: 
the CombMAX method, which performed well on the expert finding setting, 
was significantly worse than the baseline methods on the blog distillation setting\cite{macdblog}.
Therefore, how will these methods perform on the thread retrieval task is the focus of this study. 

Several combination techniques have been proposed to address evidence combination for thread retrieval. 
\cite{elsas2009} proposed two strategies to rank threads: inclusive
and selective. The inclusive strategy utilizes evidence from all messages in order
to rank parent threads. Two models from previous work on blog site retrieval
\cite{elsas2008} were adapted to thread search: the large document and the small document
models. The large document model creates a virtual document for each thread
by concatenating the thread's message texts, then it scores threads based on their
virtual document relevance to the query. In contrast, the small document model
defines a thread as a collection of text units (messages). Then, it scores threads
by adding up their messages’ relevance scores.
In contrast to the inclusive strategy, \cite{elsas2009}'s selective strategy treats threads as collections of messages; 
and it uses only few messages to rank threads. 
Three selective methods were used. The first one is scoring threads using only the
initial message relevance score. The second method scores threads by taking the
maximum score of their message relevance scores. The third method is based on
the Pseudo Cluster Selection(PCS) method\cite{seo2008}. 
PCS scores threads in two steps:
it scores a list of messages, then it ranks threads by taking the geometric mean of
the top $k$ ranked messages' scores from each thread. Generally, it was found that
the selective models are statistically superior to the inclusive models\cite{elsas2009,elsas2011}. 
Our work extends this selective strategy by investigating more aggregation methods.
In addition, PCS focuses on the top $k$ ranked messages, whereas we focus on all ranked messages. 
Applying voting techniques as aggregation methods in PCS is an interesting problem, but we leave it for a separate study.

Another line of research is the multiple context retrieval approach proposed
by \cite{seo2011}. 
This approach treats a thread as a collection of several local contexts---
types of self-contained text units. 
Four contexts were proposed: posts--- identical to messages, pairs, dialogues and the entire thread. The thread and post contexts
are identical to \cite{elsas2009}'s virtual document and message based representations.
In the pair and the dialogue contexts, the conversational relationship between
messages is exploited to build text units. In the pair context, for each pair
of messages $m_i,m_j$
that have a reply relationship--- $m_j$ is a reply to $m_i$, a text unit is built by concatenating their texts. 
In the dialogue context, for each chain of replies that starts by the initial message;
and, there is a reply relation between each message and its neighbour in the chain, a text unit is built by
concatenating the chain's message texts. To rank threads using the post, pair
and dialogue contexts, PCS was used. It was observed that the retrieval using the
dialogue context outperformed retrieval using other contexts. Additionally, the
weighted product between the thread context and the dialogue contexts achieved
the best performance. In our work, we are focusing on how to combine the ranked
contexts' relevance scores. 
Therefore, our work is complementary to \cite{seo2011}'s work.

The third line of work is the structure based document retrieval proposed by\cite{bhatia2010}. 
In this approach, a thread consists of a collection of structural components:
the title, the initial message and the reply messages set. 
In this representation, the thread relevance to the user query is estimated using \cite{metzler2004}'s inference network
framework. 
Our work can be applied to \cite{bhatia2010}'s representation as well. We could
use \cite{bhatia2010}'s inference based relevance score the same way the thread context score was used in \cite{seo2011}.

\section{Experimental Design}
Thread retrieval is a new task, and the number of test collections is limited.
In this study, we used the same corpus used by \cite{bhatia2010}. 
It has two datasets from two forums--- Ubuntu\footnote{ubuntuforums.org} and
Travel\footnote{http://www.tripadvisor.com/ShowForum-g28953-i4-New York.html}
forums.
The statistics of the corpus is given in Table \ref{table:dataset}.
Text was stemmed with the Porter stemmer, and stopword removal was applied at the ranking stage.
In conducting the experiments, we used the Indri retrieval
system\footnote{http://www.lemurproject.org/indri.php}.

\begin{table}
\begin{center}
\caption{Statistics of test collection}
\label{table:dataset} 
\begin{tabular}{lcc}
\hline
 & Ubuntu  & Travel \tabularnewline\\\hline
No of threads  & 113277  & 83072\tabularnewline
No of users  & 103280  & 39454\tabularnewline
No of messages  & 676777  & 590021\tabularnewline
No of queries  & 25  & 25\tabularnewline
No of judged threads  & 4512  & 4478\tabularnewline
\hline
\end{tabular}
\end{center}
\end{table}
As for evaluation, we use \cite{elsas2009}'s virtual document model $VD$ as a baseline. 
This model has been used as a strong baseline in previous studies \cite{elsas2009,seo2011,bhatia2010}. 
For each query, we calculated the standard used measures on Ad Hoc retrieval \cite{mark2010}: 
Precision at 10 (P@10), Normalized Discounted Cumulative Gain at 10 (NDCG@10),
Mean Reciprocal Rank(MRR) and Mean Average Precision (MAP). 
In all experiments, we used the same relevance protocol followed in \cite{bhatia2010,seo2011}, a thread is
considered as relevant if its relevance level is greater or equal to 1--- if it is
partially or highly relevant; and, it is irrelevant if the relevance level is zero.

As for parameter estimation, we estimated the smoothing parameters, $\mu$, for
the virtual document and message language models. 
In addition, for all voting techniques, we estimated the size of the initial ranked list of messages $R_Q$. 
To estimate $\mu$, we varied its value from 500 up to 4000; adding 500 in each run. 
To estimate the size of $R_Q$, we varied its value from 500 up to 5000 adding 500 in each run.
Then, an exhaustive grid search was applied to maximize MAP using 5-fold cross validation.

\section{Result and Discussion}

\begin{table}
\begin{center}
\caption{Retrieval performance of the voting methods on the Ubuntu dataset}
\label{tbl-ra-navie-u}

\begin{tabular}{lllll}
\hline
Method      & MAP         & MRR         & P@10 & NDCG@10\\\hline
VD          & 0.3437$^{}$ & 0.7258$^{}$ &  0.4200$^{}$ &  0.3284$^{}$  \\ 
CombGNZ     & 0.2272$^{\triangledown}$ &  0.4974$^{\blacktriangledown}$ &  0.2760$^{\triangledown}$ &  0.1971$^{\triangledown}$  \\ 
Votes       & 0.2749$^{\triangledown}$ &  0.6550$^{}$ &  0.4680$^{}$ &  0.3551$^{}$  \\
RR          & 0.3313$^{}$ &  0.6287$^{}$ &  0.4600$^{\blacktriangle}$ &  0.3428$^{}$ \\
Bordafuse   & 0.3153$^{}$ &  0.6913$^{}$ &  0.5080$^{}$ &  0.3778$^{}$  \\
CombMIN     & 0.1779$^{\triangledown}$ &  0.5000$^{\blacktriangledown}$ &  0.2600$^{\triangledown}$ &  0.1849$^{\triangledown}$  \\
CombMAX     & 0.3074$^{\triangledown}$ &  0.6420$^{}$ &  0.4480$^{}$ &  0.3257$^{}$  \\
CombMED     & 0.2212$^{\triangledown}$ &  0.5021$^{\blacktriangledown}$ &  0.2760$^{\triangledown}$ &  0.1927$^{\triangledown}$  \\
CombSUM     & 0.3100$^{}$ &  0.6667$^{}$ &  0.4720$^{}$ &  0.3633$^{}$  \\
CombANZ     & 0.2314$^{\triangledown}$ &  0.4971$^{\blacktriangledown}$ &  0.2800$^{\triangledown}$ &  0.1991$^{\triangledown}$  \\
CombMNZ     & 0.3108$^{}$ &  0.6933$^{}$ &  0.4880$^{}$ &  0.3720$^{}$  \\
expCombSUM  & 0.3088$^{}$ &  0.6933$^{}$ &  0.4840$^{}$ &  0.3676$^{}$  \\
expCombANZ  & 0.2315$^{\triangledown}$ &  0.4971$^{\blacktriangledown}$ &  0.2800 $^{\triangledown}$ & 0.1991$^{\triangledown}$  \\
expCombMNZ  & 0.3088$^{}$ &  0.6933$^{}$ &  0.4840$^{}$ &  0.3676$^{}$  \\\hline
\end{tabular}
\end{center}
The symbols $^{\triangle}$ and  $^{\blacktriangle}$ denote statistically significant improvements over the virtual document model (VD) at p-value $<$ 0.01 and 0.05 respectively using paired randomization test. 
Similarly, $^{\triangledown}$ and  $^{\blacktriangledown}$ denote  statistically significant degradations over (VD) at p-value $<$ 0.01 and 0.05 respectively.
\end{table}

\begin{table}
\begin{center}
\caption{Retrieval performance of the voting methods on the Travel dataset}
\label{tbl-ra-navie-t}

\begin{tabular}{lllll}
\hline
Method & MAP & MRR & P@10 & NDCG@10 \\\hline
VD & 0.3774 & 0.6967 & 0.4800 & 0.3549\\
CombGNZ & 
0.2001$^{\triangledown}$ & 0.4838$^{\blacktriangledown}$ & 0.3320$^{\triangledown}$ & 0.2319$^{\triangledown}$ \\
Votes &
0.3066$^{\triangledown}$ & 0.7491$^{}$ & 0.5080$^{}$ & 0.4063$^{}$ \\
RR & 
0.3155$^{\triangledown}$ & 0.6120$^{}$ & 0.4520$^{}$ & 0.3431$^{}$ \\
BordaFuse & 
0.3630$^{}$ & 0.7547$^{}$ & 0.5640$^{}$ & 0.4350\\
CombMIN & 
0.1574$^{\triangledown}$ & 0.4843$^{\blacktriangledown}$ & 0.3040$^{\triangledown}$ & 0.2199$^{\triangledown}$ \\
CombMAX & 
0.2724$^{\triangledown}$ & 0.5754$^{}$ & 0.4360$^{}$ & 0.3216$^{}$ \\
CombMED &
0.2004$^{\triangledown}$ & 0.4841$^{\blacktriangledown}$ & 0.3480$^{\triangledown}$ & 0.2388$^{\triangledown}$ \\
CombSUM &
0.3668$^{}$ & 0.8000$^{}$ & 0.5560$^{}$ & 0.4440$^{\blacktriangle}$ \\
CombANZ & 
0.2065$^{\triangledown}$ & 0.4841$^{\blacktriangledown}$ & 0.3400$^{\triangledown}$ & 0.2346$^{\triangledown}$\\
CombMNZ & 
0.3575$^{}$ & 0.7790$^{}$ & 0.5280$^{}$ & 0.4205$^{}$ \\
expCombSUM & 
0.3513$^{}$ & 0.7937$^{}$ & 0.5200$^{}$ & 0.4109$^{}$ \\
expCombANZ & 
0.2065$^{\triangledown}$ & 0.4841$^{\blacktriangledown}$ & 0.3400$^{\triangledown}$ & 0.2346$^{\triangledown}$ \\
expCombMNZ &
0.3513$^{}$ & 0.7937$^{}$ & 0.5200$^{}$ & 0.4109$^{}$\\\hline
\end{tabular}
\end{center}
The symbols $^{\triangle}$ and  $^{\blacktriangle}$ denote statistically significant improvements over the virtual document model (VD) at p-value $<$ 0.01 and 0.05 respectively using paired randomization test. 
Similarly, $^{\triangledown}$ and  $^{\blacktriangledown}$ denote  statistically significant degradations over (VD) at p-value $<$ 0.01 and 0.05 respectively.
\end{table}

Table \ref{tbl-ra-navie-u} and Table \ref{tbl-ra-navie-t} present the retrieval performance of the voting methods on
thread retrieval for the Ubuntu dataset and the Travel dataset respectively.
Several observations can be found from the data shown in these tables.
The first observation is the performance of the aggregation methods as compared to the “baseline method”--- 
the virtual document(VD) model. In high precision measures (P@10 and NDCG@10),
RR, BordaFuse, CombSUM, CombMNZ, expCombSUM, expCombSUM and expCombMNZ are able to produce better or
comparable result with respect to VD. These methods favour threads with highly
ranked messages. In contrast, CombGNZ, CombMED, CombANZ, CombMIN
and expCombANZ might be effected by threads that have a lot of low scored
messages. This behaviour was also reported in applying voting techniques to expert finding[6]. 
Therefore, based on \cite{macdexpert}'s conclusion, we assert that highly ranked messages are good indicators of relevant threads.

\begin{figure}[h!]
   \centering
   \includegraphics[width=\textwidth]{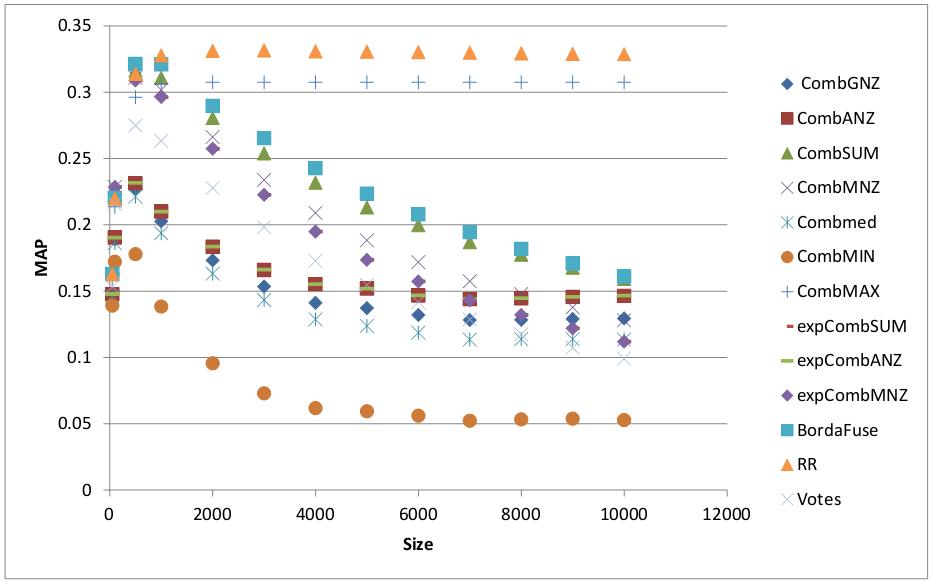} 
   \caption{The performance of aggregation methods as the size of the initial ranked list increases on the Ubuntu dataset.}
  \label{fig-u-ra-navie-size-map}
          
\end{figure}

\begin{figure}[h!]
  \centering
   \includegraphics[width=\textwidth]{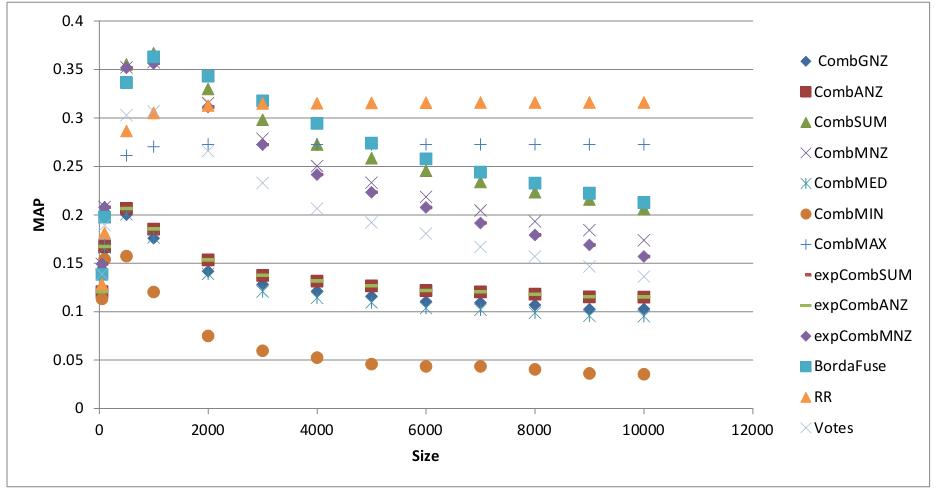}
  \caption{The performance of aggregation methods as the size of the initial ranked list increases on the Travel dataset.}
  \label{fig-t-ra-navie-size-map}       
\end{figure}
To confirm this conclusion, the effects of varying the size of the initial ranked
list was studied. As Figure \ref{fig-u-ra-navie-size-map} and Figure \ref{fig-t-ra-navie-size-map} show, the retrieval performance decreases as the
size gets relatively big (more than 1000). In addition, one can see that almost all
methods suffer from this problem except RR and CombMAX. This is expected
because RR and CombMAX address the problem of low scored messages inherently. RR adds up the inverse of the messages' ranks, 
thus it penalizes threads with a lot of low ranked messages.
In the case of CombMAX, it takes only the best scoring message; therefore, if no threads are introduced as the size increases,
the order of threads will not change. That explains the convergence of CombMAX and RR and the consistent decrement of the other methods. 
This was replicated with other measures such as P@10 and NDCG@10(Not shown in this paper) as well.
This indicates the importance of highly ranked messages to thread retrieval.

Another observation is the importance of utilizing non score signals. 
For instance, the Votes method's performance is relatively good as compare to other
methods. Similarly, CombMNZ, which makes use of the number of ranked messages in addition to sum of scores, 
has similar performance as well. 
All of these methods leverage information that is not coming from scores: the number of
ranked messages. Nevertheless, exhaustive emphasis on these signals will hurt the
performance. One could see that from fast decrement of Votes and CombMNZ
methods as size increases. One possible reason is that adding up low scores
has less impact than multiplying by the number of these messages; CombSUM's decrement is always less than those of the Votes and the CombMNZ methods.

Although the voting methods improvements are not statistically significant,
they are consistent on both datasets and require only using the message index.
That gives the voting approach an extra advantage over the virtual document
model because it coincides with what users contribute, hence it frees the retrieval
system from re-concatenating messages into a virtual document whenever a new
message is created or edited.

\section{Conclusion}
In this paper, we studied applying voting techniques to online forums thread retrieval. 
We used thirteen voting methods that aggregate ranked messages’ scores
or ranks in order to score the parent threads. The experimental result shows that
voting techniques---RR, BordaFuse, CombSUM, CombMNZ, expCombSUM, expCombSUM and expCombMNZ, that favour threads with highly ranked messages produced comparable or better performance as compare to baselines; and, none of them is a winning method. Although the observed improvements were
not statistically significant, we recommend using the voting methods because their
improvements are consistent across datasets, and they coincide with what users contribute.

Nevertheless, this paper finding has motivated us to further study the effects of voting techniques when aggregating only the top $k$ messages.
Another future direction is incorporating these voting methods into \cite{seo2011}'s
multiple context models. Similar approach will be applied to incorporate the
structural component representation of \cite{bhatia2010}.

\subsubsection*{Acknowledgements} 
This work is supported by Ministry of Higher Education (MOHE) and Research Management Centre (RMC) at the Universiti Teknologi Malaysia (UTM) under Research University Grant Category (VOT Q. J130000. 7128. 00H72).

\end{document}